\documentclass[rsi,twocolumn]{revtex4}  

\usepackage{graphicx}   
\usepackage{dcolumn}    
\usepackage{bm}         
\usepackage{amssymb}    
\usepackage{amsmath}    
\usepackage{tabularx}
\usepackage{textcomp}
\usepackage{placeins}
\usepackage[pdfborderstyle={/S/U/W 1}]{hyperref}
\usepackage{color}
\usepackage{verbatim}

\def\H        {{$^1$H \/}}

\def\C        {{$^{13}$C \/}}

\def\NN       {{$^{15}$N \/}}

\newcommand{\ee}[1]{\times 10^{#1}}
\newcommand{\mr}[1]{\mathrm{#1}}
\newcommand{\unit}[1]{\,\mathrm{#1}}
\newcommand{\uH}{\,\mu{\rm H}}
\newcommand{\um}{\,\mu{\rm m}}
\newcommand{\us}{\,\mu{\rm s}}

\newcommand{\braket}[1]{\ensuremath{\left\langle#1\right\rangle}}

\newcommand{\Bpar}{B_{||}}
\newcommand{\Bperp}{B_{\perp}}
\newcommand{\df}{\delta f}
\newcommand{\dw}{d_\mr{w}}
\newcommand{\fdb}{f_\mr{3dB}}
\newcommand{\frf}{f_\mr{RF}}
\newcommand{\ms}{m_S}
\newcommand{\tscan}{t_\mr{scan}}
\newcommand{\rhow}{\rho_\mr{w}}
\newcommand{\ye}{\gamma_\mr{e}}

\begin{document}

\title{High-bandwidth microcoil for fast nuclear spin control}

\author{K. Herb}
\email{science@rashbw.de}
\author{J. Zopes}
\author{K. S. Cujia}
\author{C. L. Degen}
\email{degenc@ethz.ch}
\affiliation{Department of Physics, ETH Zurich, Otto Stern Weg 1, 8093 Zurich, Switzerland}
\date{\today}

\begin{abstract}
The active manipulation of nuclear spins with radio-frequency (RF) coils is at the heart of nuclear magnetic resonance (NMR) spectroscopy and spin-based quantum devices.  Here, we present a microcoil transmitter system designed to generate strong RF pulses over a broad bandwidth, allowing for fast spin rotations on arbitrary nuclear species.  Our design incorporates: (i) a planar multilayer geometry that generates a large field of 4.35\,mT per unit current, (ii) a 50\,$\Omega$ transmission circuit with a broad excitation bandwidth of approximately 20\,MHz, and (iii) an optimized thermal management for removal of Joule heating.  Using individual $^{13}$C nuclear spins in the vicinity of a diamond nitrogen-vacancy (NV) center as a test system, we demonstrate Rabi frequencies exceeding 70\,kHz and nuclear $\pi/2$ rotations within 3.4\,$\mu$s.  The extrapolated values for $^{1}$H spins are about 240\,kHz and 1\,$\mu$s, respectively.  Beyond enabling fast nuclear spin manipulations, our microcoil system is ideally suited for the incorporation of advanced pulse sequences into micro- and nanoscale NMR detectors operating at low ($<$1\,T) magnetic field.
\end{abstract}

\maketitle

\section{Introduction}

The active control of nuclear spins in the form of strong radio-frequency pulses is a common method in conventional NMR to suppress line broadening \cite{waugh68} and for advanced multi-dimensional spectroscopy \cite{aue76}.
High-field NMR uses tuned circuits to achieve large radio-frequency fields while simultaneously maximizing the detection sensitivity. 
By shrinking coils to micrometer dimensions and driving them with kilowatt amplifiers, proton Rabi frequencies exceeding $1\unit{MHz}$ have been demonstrated \cite{yamauchi04,webb13}.

In recent years several applications have emerged that operate at low magnetic field and that require broadband excitation, calling for alternative radio-frequency transmitter circuits.  One promising application are micro- and nanoscale NMR detectors based on nitrogen-vacancy (NV) impurities in diamond, which aim at detecting NMR signals originating from molecules near the surface of a diamond chip  \cite{staudacher13,mamin13}.  At the microscale, NV-NMR detectors are expected to provide a new route to microfluidic analysis of small sample quantities \cite{glenn18,smits19}.  If successfully scaled down to the single-molecule level, NV-NMR spectroscopy would add the capability for direct imaging of three-dimensional molecular structures \cite{ajoy15,wrachtrup16,schwartz17}.  Another important area for nanoscale nuclear spin control are multi-qubit spin registers \cite{dutt07,pla13,taminiau14,bradley19} and quantum memories \cite{morton08,pfender17,rosskopf17} demanding efficient quantum gate operations.
All of these applications benefit from strong RF fields, requiring circuits that can accept and dissipate high RF powers \cite{lovchinsky16}.

Here, we demonstrate a microcoil excitation system designed for the generation of strong, time-varying magnetic fields at frequencies up to $20\unit{MHz}$.
Our untuned, broadband circuit allows actuating a broad range of nuclear species, including protons, at fields up to several hundred millitesla.  Coil designs for larger bandwidth or higher field per unit current are also presented.  The coil holder structure is optimized for efficient heat extraction such that large currents up to several $A$ can be applied.  We calibrate the magnitude and transient response of the coil magnetic field \textit{in situ} using optically-detected magnetic resonance (ODMR) spectroscopy of an NV center in diamond.  We demonstrate the coil functionality by driving fast Rabi rotations of a single \C nuclear spin in the NV center's vicinity.


\section{Implementation}

\subsection{Experimental setup}
\begin{figure}
	\includegraphics[width=0.42\textwidth]{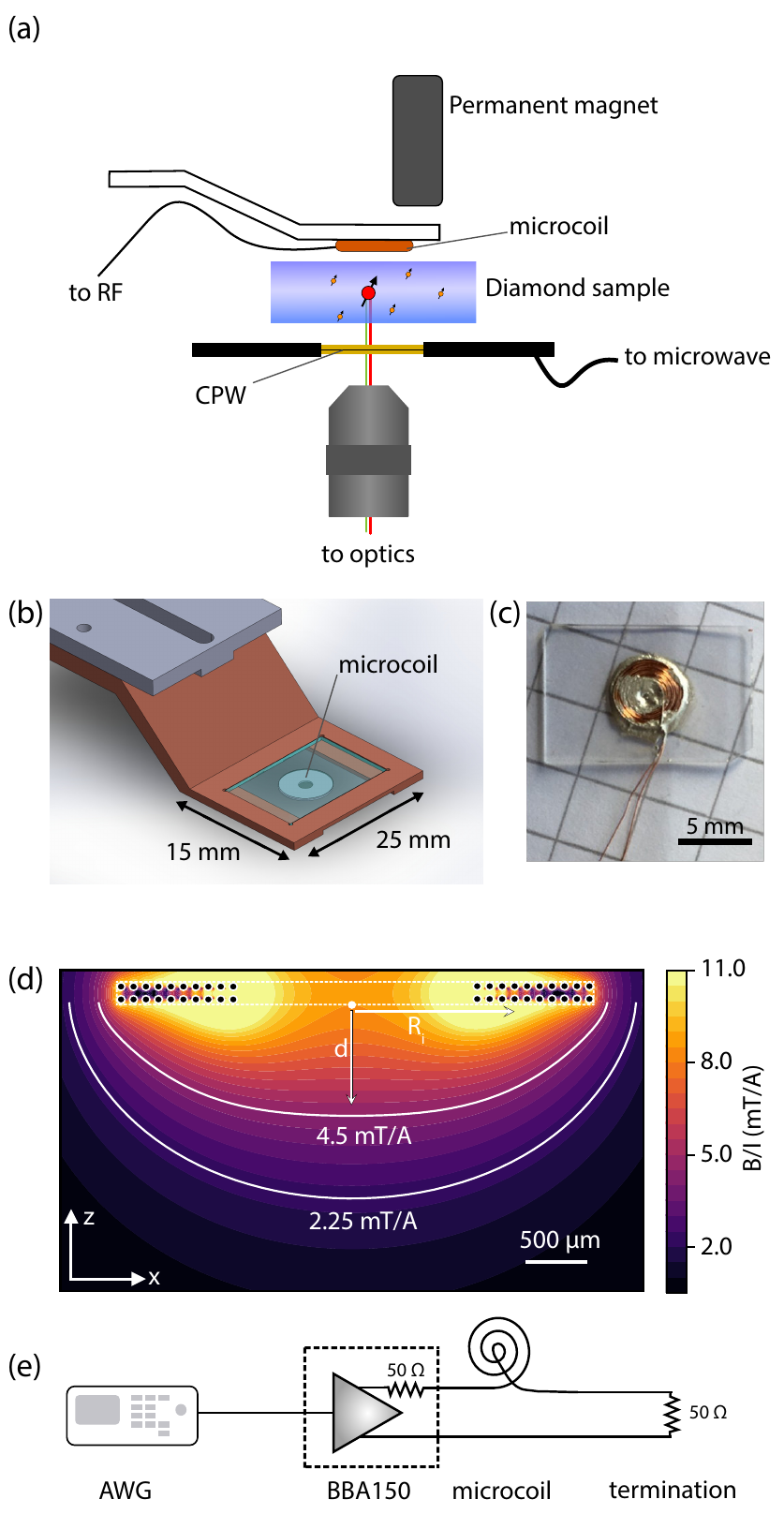}
	\caption{
	Experimental setup and microcoil design.
	(a) Schematic of the ODMR setup as described in the text (not to scale). The NV center is indicated by the arrow-crossed red circle, the adjacent $^{13}\mathrm{C}$ nuclear spins by the smaller arrow-crossed circles.
	(b) Mechanical mount of the coil, including coil (central disk), mounting plate (transparent blue), copper holder (brown) and aluminum bracket (gray).  The mechanical mount is attached to a three-axis translation stage controlled by manual micrometer screws (not shown).
	(c) {Photograph of microcoil design \textnumero1 glued to a CVD diamond plate}.  Microcoils are produced by Sibatron (Switzerland) and consists of 100-$\um$-thick copper magnet wire wound in the shape of an Archimedean spiral.  The wires are isolated with a 20-$\um$-thick layer of varnish. 
	(d) Magnetic field distribution of the coil, calculated by numerically evaluating Biot and Savart's law and summing over loop currents.  Two isofield contours at $B=4.5\unit{mT/A}$ and $B=2.25\unit{mT/A}$ are shown.  {Shown is the vertical cross section ($xz$ plane).  $d$ is the vertical distance between nuclear spins and the lower coil surface.}
	(e) Radio-frequency drive circuit.  Pulses are generated by direct synthesis on an arbitrary waveform generator (AWG, NI PCI-5421) and amplified by a broadband amplifier (Rohde \& Schwarz BBA150). A $50\unit{\Omega}$ termination in series with the coil is used to match the impedance of the amplifier and to reduce the quality factor.
	}
	\label{fig:setup_drawing}
\end{figure}

Fig. \ref{fig:setup_drawing}(a) shows a schematic of the ODMR setup into which we integrate the microcoil system.  From bottom to top, the setup includes a microscope objective for optical initialization and readout of the NV center, a coplanar waveguide (CPW) to drive the NV center's $\sim 2.9\unit{GHz}$ electronic spin transition, the diamond crystal containing the NV center and \C nuclei, the microcoil, and a permanent magnet for applying a bias field.  The microcoil is positioned via a mechanical mount that can be translated in three spatial directions.  

\subsection{Microcoil design}

Our goal is to design a microcoil transmitter circuit that produces a large RF field over a bandwidth sufficient for addressing a broad range of nuclear species with resonance frequencies up to tens of MHz.  Further design constraints are a sufficient vertical clearance between the coil surface and the sample and a maximum current that can be applied without overheating the circuit.

Before describing our implementation, we recall the basic design parameters underlying microcoil transmitters.  For a broadband RF circuit with no resonant tuning, the optimum coil geometry is a trade-off between bandwidth and RF field amplitude.  To achieve a large bandwidth, the coil must have a low inductance $L$, which in turn requires a low number of coil windings.  On the other hand, the more windings, the larger the magnetic field $B$ generated per unit current $I$.  The goal therefore is to find a geometry that adequately balances between bandwidth and field strength.

We estimate the expected magnetic field and inductance of a multilayer solenoid by summing over $N$ circular loop currents,
\begin{align}
B &\approx \sum\limits_{i=1}^{N} \frac{\mu_0 I R_i^2}{2([z_i-d]^2+R_i^2)^{3/2}}
   \approx \frac{\mu_0 I N \bar{R}^2}{2([h/2-d]^2+\bar{R}^2)^{3/2}}  \label{eq:B} \\
L &\approx {\mu_0} N^2 \bar{R} \, G_1 G_2 \label{eq:L}
\end{align}
where $R_i$ is the radius and $z_i$ is the vertical position of the $i$'th loop, referenced to the bottom of the coil (see Fig. \ref{fig:setup_drawing}(d)). $\bar{R}$ is the average radius of the coil windings and $h$ is the coil height.  $d$ is the vertical distance between the lower coil surface and the spins' location in the sample, and $N = (\text{number of windings})\times(\text{number of layers})$. $G_1$ and $G_2$ are dimensions-less correction factors that depend on the coil geometry (see Appendix A).  Eqs. (\ref{eq:B}) and (\ref{eq:L}) indicate that the magnetic field and inductance approximately scale as $B\propto N$ and $L\propto N^2$, respectively.

\begin{figure}
	\includegraphics[width=0.46\textwidth]{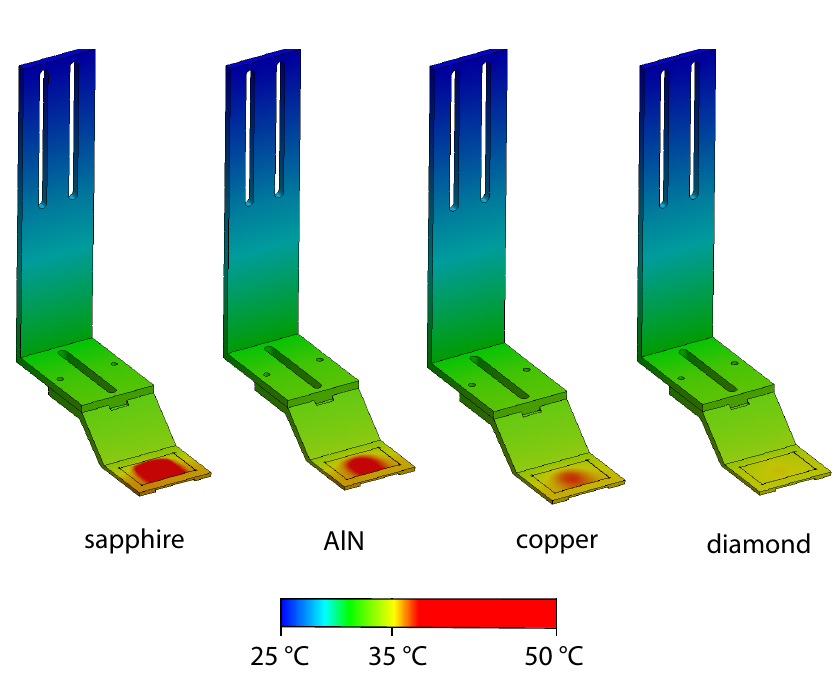}
	\caption{Finite element simulation of the temperature distribution in the coil mount, for four different mounting plate materials.  A heat load of $1.35\unit{W}$ is applied at the coil position, while the top of the back side of the aluminum bracket is thermally anchored at $25\unit{^\circ C}$.}
\label{fig:fem_sim}
\end{figure}

\begin{table*}
	\begin{center}
		\renewcommand{\arraystretch}{1.2}
		\begin{tabularx}{1.00\textwidth}{X|X|X|X|X|X|X|X|X|X}
		\hline\hline
		coil design           & layers     & windings    & ID (mm)       & OD (mm)       & $h$ (mm)      & $B/I$ (mT/A) & $L$ ($\uH$)   & $\fdb$ (MHz)  & $P$ (W)        \\
		\hline
		\textbf{\textnumero1} & \textbf{2} & \textbf{10} & \textbf{1.02} & \textbf{3.13} & \textbf{0.24} & \textbf{4.5} & \textbf{0.77} & \textbf{20.7} & \textbf{0.38}  \\
		\textnumero2          & 3          & 13          & 0.84          & 2.73          & 0.36          & 8.0          & 2.5           &  6.37         & 0.78           \\
		\textnumero3          & 1          & 11          & 1.30          & 3.62          & 0.12          & 2.4          & 0.31          & 51.3          & 0.25           \\
		\textnumero4          & 3          &  9          & 2.00          & 3.90          & 0.36          & 5.2          & 2.5           &  6.37         & 0.73           \\
		\hline\hline
		\end{tabularx}
	\end{center}
	\caption{Parameters for four coil designs. The geometries of the planar solenoids are defined by the inner diameter (ID), outer diameter (OD), height $h$, the number of layers, and the number of windings.  ID values are restricted to the diameters offered by the manufacturer (Sibatron). 
	The magnetic field $B$ and the inductance $L$ are calculated based on the geometric parameters of the coil, as explained with Eqs. (\ref{eq:B},\ref{eq:L}).  $B$ is the on-axis field at $d=1\unit{mm}$ below the coil.
  The -3-dB bandwidth $\fdb$ is calculated using Eq. (\ref{eq:fdb}).
	The dissipated power $P$ is calculated using Eq. (\ref{eq:P}) for an rms current of $I=1\unit{A}$, a wire resistance of $\rho = 1.7 \cdot 10^{-8}\unit{\Omega/m}$ and by setting the pulse frequency to $2\unit{MHz}$ to estimate the skin effect.  The skin effect causes a increase of the resistance by a factor of 1.39 (see Appendix A).
  Most reported experiments are performed with design \textnumero1 (bold).
	}
	\label{tab:setup_coil_properties}
\end{table*}

The $-3\unit{dB}$ bandwidth $\fdb$ of the coil and the dissipated power $P$ are given by
\begin{align}
\fdb &= 100\unit{\Omega}/(2\pi L) \label{eq:fdb} \ , \\
P &\approx \sum\limits_{i=1}^{N} 2\pi R_i \alpha \rhow I^2  \ , \label{eq:P}
\end{align}
where the $100\unit{\Omega}$ represent the sum of the terminal resistance and the output impedance of the amplifier (see Fig. \ref{fig:setup_drawing}e).  $\rhow=1.7\ee{-8}\unit{\Omega/m}$ is the DC resistance of the 100-$\um$-thick magnet wire and $\alpha$ is a frequency-dependent correction factor accounting for the skin effect (see Appendix A).
To find an optimum coil configuration, we vary the coil geometry such that $B/I$ is maximized under the given design constraints for $\fdb$ and the vertical distance $d$. We use a wire thickness of $100\unit{\um}$ in all designs.

In Table \ref{tab:setup_coil_properties} we compare several coil geometries and their computed inductance and magnetic field.  For our experimental demonstration, we choose design \textnumero1 that is optimized for a bandwidth of $\fdb=20\unit{MHz}$ and a distance $d=1\unit{mm}$, and provides approximately $4.5\unit{mT}$ field per unit current.
Design \textnumero2 provides higher field at reduced bandwidth, while design \textnumero3 provides a larger bandwidth but lower field.  Design \textnumero4 allows for a larger working distance $d$ and has a larger tolerance in the mechanical alignment.
A photograph of the microcoil \textnumero1 is shown in Fig. \ref{fig:setup_drawing}(c) and the two-dimensional field distribution is shown in Fig. \ref{fig:setup_drawing}(d).

\subsection{Thermal anchoring}

To maximize the current that can be applied to the circuit, we thermally anchor the coil on a holder structure that is optimized for efficient heat extraction.
A schematic of the holder assembly is shown in Fig. \ref{fig:setup_drawing}(b).  The coil is glued by a high-thermal-conductivity epoxy to a $10\unit{mm}\times 15\unit{mm}$ mounting plate  that sits on a larger copper holder.  The copper holder is connected to a large aluminum bracket that acts as the terminal heat sink.  The coil feed lines are soldered to a small printed circuit board and routed to SMA sockets.


\section{Characterization}

\subsection{Thermal characterization}

The most critical heat links are the thermal contact between coil and mounting plate, as well as the plate material. We investigate four plate substrates: sapphire (EMATAG AG Switzerland), aluminium nitride (CeramTec Germany), copper, and CVD diamond (Diamond Materials Germany).  We further investigate two epoxy resins for coil attachment: Epotek H20E and Masterbond Supreme 18TC.
To choose a plate substrate, we use a finite element software (Solidworks) to simulate the equilibrium temperature distribution in the coil holder assembly while applying a thermal load of $1.35\unit{W}$ at the coil position.  We connect the back side of the L-shaped aluminum holder to a thermal reservoir at $300\unit{K}$, and surround the entire holder structure by air at $300\unit{K}$ with a convention rate of $25\unit{W m^{-2} K^{-1}}$.

In Fig. \ref{fig:fem_sim} we show the temperature distribution for the four assemblies.  Clearly, heat removal is most efficient for the diamond substrate, thanks to its exceptional thermal conductivity of $2,300\unit{W m^{-1} K^{-1}}$.  The second-best performance results for OFHC copper, while sapphire and aluminum nitride are not competitive.  {The simulation predicts a temperature increase of $10\unit{K}$ at the coil position for the diamond substrate, and of $13\unit{K}$, $15\unit{K}$ and $26\unit{K}$ for the copper, aluminum nitride and sapphire substrates, respectively (see Fig. \ref{fig:fem_sim}).  For our particular setup, we strive to keep the temperature rise below $2\unit{K}$ in order to avoid drifts in the optical alignment and the external bias field.  This corresponds to a maximum dissipated power in the microcoil of $\sim 0.27\unit{W}$.  The maximum temperature tolerated by the coil assembly before suffering structural damage is $>370\unit{K}$.}

\subsection{Electrical characterization}
\label{sec:elec_char}

\begin{figure}[t!]
	\includegraphics[width=0.47\textwidth]{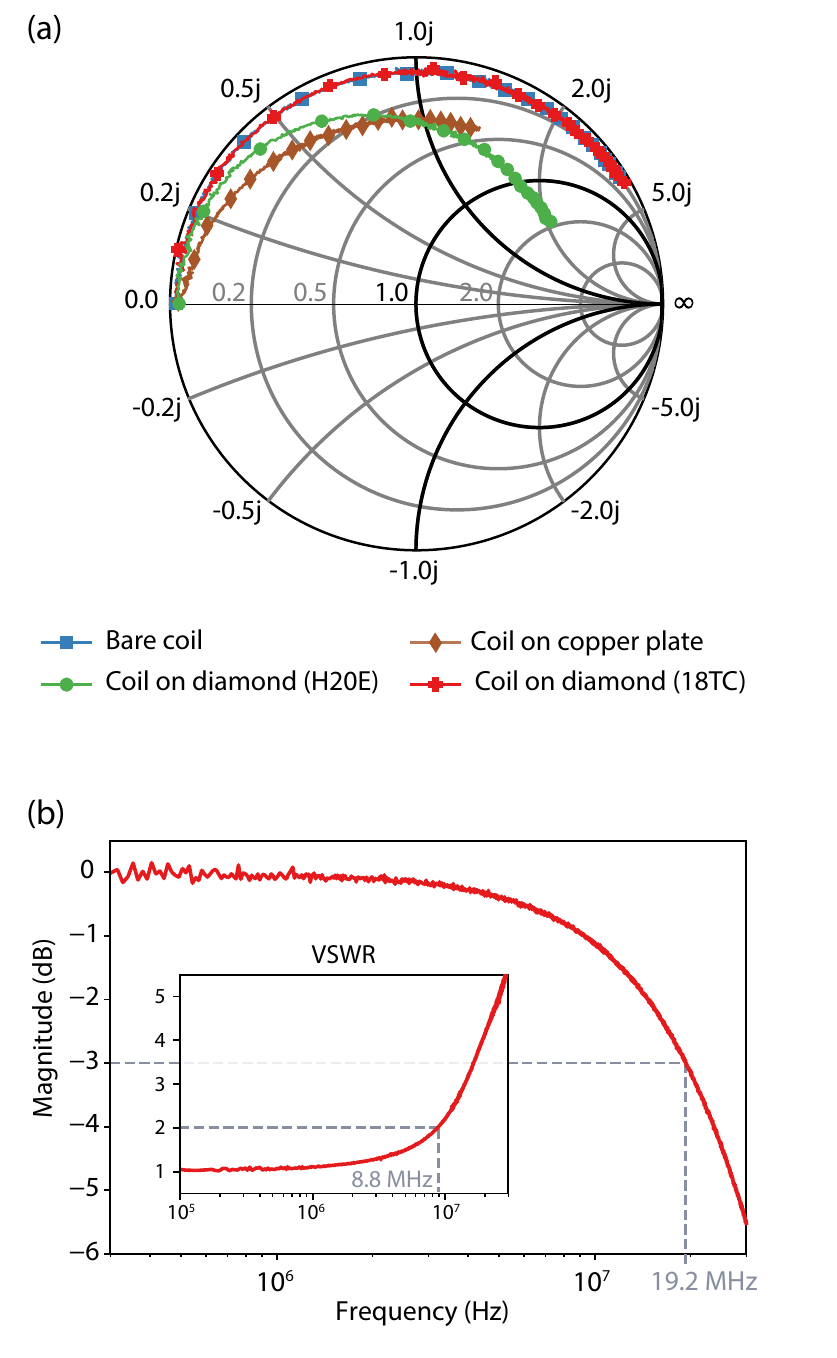}
	\caption{
	Electrical characterization.
(a) Smith chart of the reflection parameter $S_{11}$ for coils (design \textnumero4) mounted on different mounting plate materials (copper, diamond) with different epoxy glues (electrically conducting, non-conducting). 
Blue data are for a bare coil with no mounting plate.
Brown data are for a coil glued with conductive epoxy to a copper mounting plate.
Green data are for a coil glued with conductive epoxy to a diamond mounting plate.
Red data are for a coil glued with non-conductive epoxy to a diamond mounting plate.
Clearly, only the electrically isolating mount preserves the high bandwidth of the bare coil.   The input frequency range is $0.01-12.5\unit{MHz}$.
(b) Transmission parameter $S_{12}$ and voltage standing wave ratio (VSWR, inset) for the complete coil circuit (design \textnumero1) connected to the 50-$\Omega$ termination.
	}
	\label{fig:characterization}
\end{figure}

Fig. \ref{fig:characterization}(a) shows a vector network analysis of unmounted and mounted coils of design \textnumero4.
For the bare coil with no mounting plate (blue data points), the imaginary part of the reflection parameter $S_{11}$ increases linearly with frequency and the real part is almost  constant.  The inferred inductance of the coil is $L=2.37\unit{\uH}$, in good agreement with the design value calculated from the geometry ($L=2.5\unit{\uH}$, see Table \ref{tab:setup_coil_properties}).

The remaining curves in Fig. \ref{fig:characterization}(a) represent coils mounted on copper and diamond substrates.  We observe that mounting the coil onto a conductive copper substrate (brown data points) creates a magnetic short of the circuit, leading to an undesired, strong reduction of the bandwidth and a reduced magnetic field strength.  A similar reduction in bandwidth occurs when mounting the coil by conductive silver epoxy (Epotek H20E) onto the diamond substrate (green data points).  Only when also replacing the silver epoxy by a non-conductive resin (Masterbond Supreme 18TC) the electrical characteristics of the bare coil are recovered (red data points).

In Fig. \ref{fig:characterization}(b), we show the measured transmission parameter $S_{21}$ and voltage standing wave ratio (VSWR) for the final coil assembly (design \textnumero1) with diamond as the mounting substrate and the insulating epoxy.  The coil is terminated with a $50\unit{\Omega}$ resistance.  We measure a $\fdb$ cut-off frequency of $19.3\unit{MHz}$, in good agreement with our design specifications ($\fdb=20.7\unit{MHz}$, see Table \ref{tab:setup_coil_properties}).  A VSWR below 2:1 is maintained up to $8.8\unit{MHz}$.

\subsection{\textit{In-situ} calibration using NV center magnetometry}

\begin{figure}[]
	\includegraphics[width=0.47\textwidth]{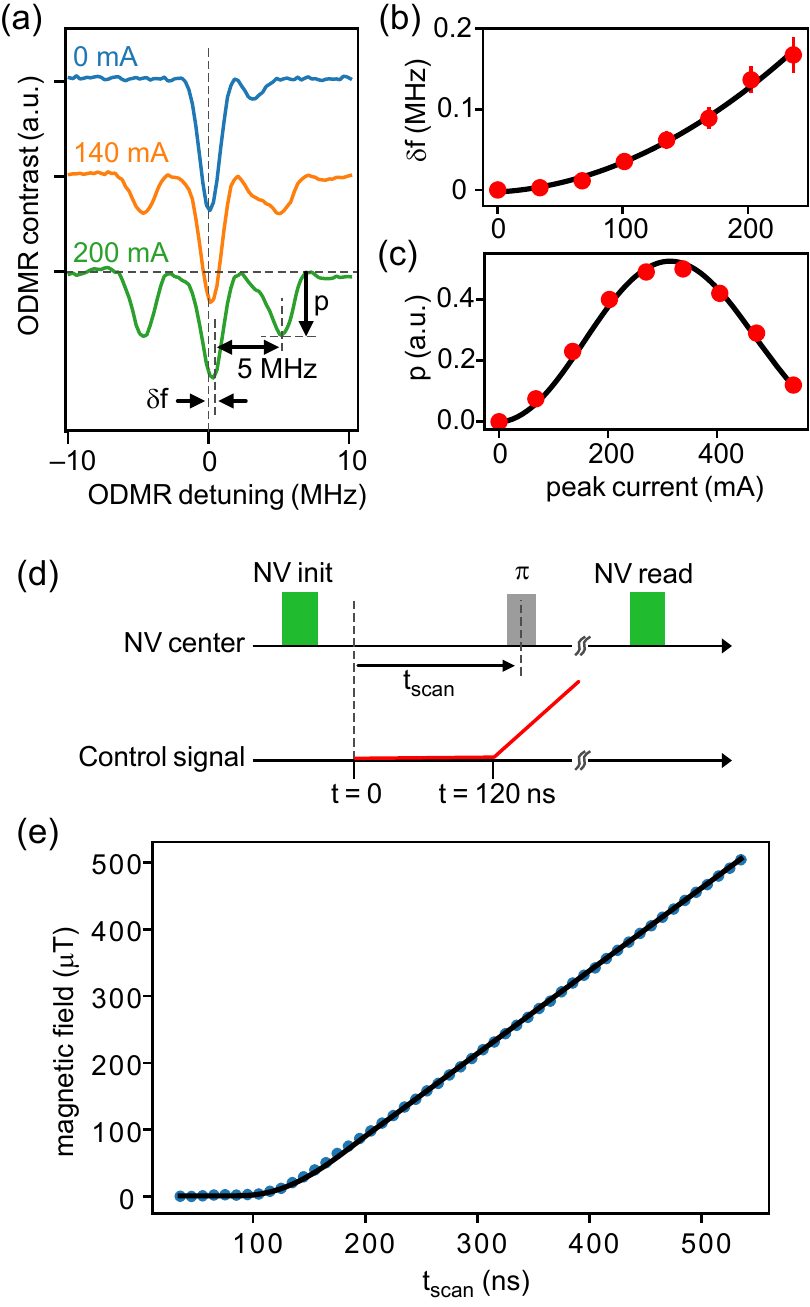}
	\caption{
	\textit{In-situ} measurement of the coil magnetic field using ODMR spectroscopy on the NV center.
	(a) ODMR spectra acquired while applying no current to the coil (upper trace) and while applying RF currents with a peak amplitude of $140\unit{mA}$ (middle trace) and $200\unit{mA}$ (lower trace).  $\df$ indicates the Bloch-Siegert shift of the center transition and $p$ the amplitude of the side band.  
	The RF frequency is $\frf=5\unit{MHz}$ and the bias field is $42\unit{mT}$.  
	Only one hyperfine transition is visible in the ODMR spectra because the \NN nuclear spin is almost fully polarized at this bias field.
	(b) Bloch-Siegert shift $\df$ as a function of applied current.  Red dots are the data and black curve a square fit, Eq. (\ref{eq:df}).
	(c) Side-band amplitude $p$ as a function of applied current. Red dots are the data and black curve a fit to a Bessel function, Eq. (\ref{eq:p}).
	(d) Timing diagram of the time-resolved ODMR measurement (green laser pulses, gray microwave pulse) and control signal applied to the coil (red curve).  The laser pulse duration is $2.5\unit{\us}$ and the $\pi$-pulse duration is $70\unit{ns}$.  We increment $t_\mathrm{scan}$ in steps of $10\unit{ns}$.
	(e) Measured coil field as a function of time (blue dots).  The black line is a fit of the exponential time response, taking the moving-average behavior of the snapshot ODMR technique into account \cite{zopes18prl}.  The fitted rise time is $8\unit{ns}$.   See Fig. \ref{fig:setup_drawing}(e) for the electrical circuit. 
	}
\label{fig:awg}
\end{figure}

We next install the microcoil system in our experimental setup and connect it to the electrical drive and dump circuits (Fig. \ref{fig:setup_drawing}(e)).  We then use pulsed ODMR spectroscopy \cite{dreau11} on the NV center to calibrate the magnitude and time response of the coil magnetic field \textit{in situ}.  Our methods are specifically selected for the calibration of AC fields.

To determine the magnitude of the field component $\Bperp$ that is perpendicular to the NV axis, we measure the Bloch-Siegert shift of the ODMR resonance under an applied AC current (Fig. \ref{fig:awg}(a)) \cite{bloch40,sacolick10}.  Since the external bias field is oriented along the NV axis, $\Bperp$ is the field component that will drive Rabi rotations of the nuclear spins.  The Bloch-Siegert shift is given by (see Appendix B):
\begin{equation}
\df \approx \frac{f_0^{(1)}+2f_0^{(2)}}{4f_0^{(1)}f_0^{(2)}} \ye^2 \Bperp^2 \ ,
\label{eq:df}
\end{equation}
where $f_0^{(1)}$ and $f_0^{(2)}$ are the frequencies of the lower (observed) and upper (unobserved) ODMR transition, $\frf$ is the RF frequency applied to the coil, $\Bperp$ is the peak field, and $\ye=28.024\unit{MHz/mT}$ is the electron gyromagnetic ratio (in units of frequency per field).  By plotting $\df$ as a function of the peak current $I$ and fitting a square law, we find the proportionality constant between $\Bperp$ and $I$ (see Fig. \ref{fig:awg}(b)).
In addition, we determine the field component parallel to the NV axis, $\Bpar$, by measuring the amplitude of the first ODMR side band appearing at $\pm \frf$ from the center peak (see Fig. \ref{fig:awg}(a)) \cite{Silveri2017}:
\begin{equation}
p \propto p_0 J_1^2\left(\frac{\ye \Bpar}{\frf}\right)  \text{.}
\label{eq:p}
\end{equation}
Here, $J_1$ denotes the first-order Bessel function of the first kind, and $p_0$ is an arbitrary and dimension-less pre-factor.
To find the proportionality constant between $\Bpar$ and $I$, we plot $p$ as a function of $I$ and fit Eq. (\ref{eq:p}) to the data (see Fig. \ref{fig:awg}(c)).
For coil design \textnumero1 we measure a parallel field of $\Bpar=0.28\unit{mT}$ and a perpendicular field of $\Bperp=0.92\unit{mT}$ at a peak current of $270\unit{mA}$.  The field magnitude extrapolated to a current of $1\unit{A}$ is $(\Bpar^2 + \Bperp^2)^{1/2}/(270\unit{mA}) = 3.6\unit{mT/A}$.  This value is slightly lower than the design value of $4.5\unit{mT/A}$ (see Table \ref{tab:setup_coil_properties}) in part due to a reduced transmission at $5\unit{MHz}$ compared to $2\unit{MHz}$, and in part due to an imprecise coil alignment.

To measure the rise time of the coil magnetic field we apply a linear-ramp input signal and record time-resolved ODMR spectra \cite{zopes18prl} by increasing the time $\tscan$ between the start trigger and ODMR pulse (Fig. \ref{fig:awg}(d)).  In this way, we can sample the temporal profile of the coil field.  (Note that a step response cannot be applied with our AC-coupled amplifier circuit).  To analyze the coil response, we fit a numerical model to the experimental data (Fig. \ref{fig:awg}(e)).  We determine a response time of $8\unit{ns}$ and a corresponding bandwidth of $19.2\unit{MHz}$, in good agreement with the vector network analysis.


\section{Nuclear Rabi rotations}

\begin{figure*}[]
	\includegraphics[width=0.8\textwidth]{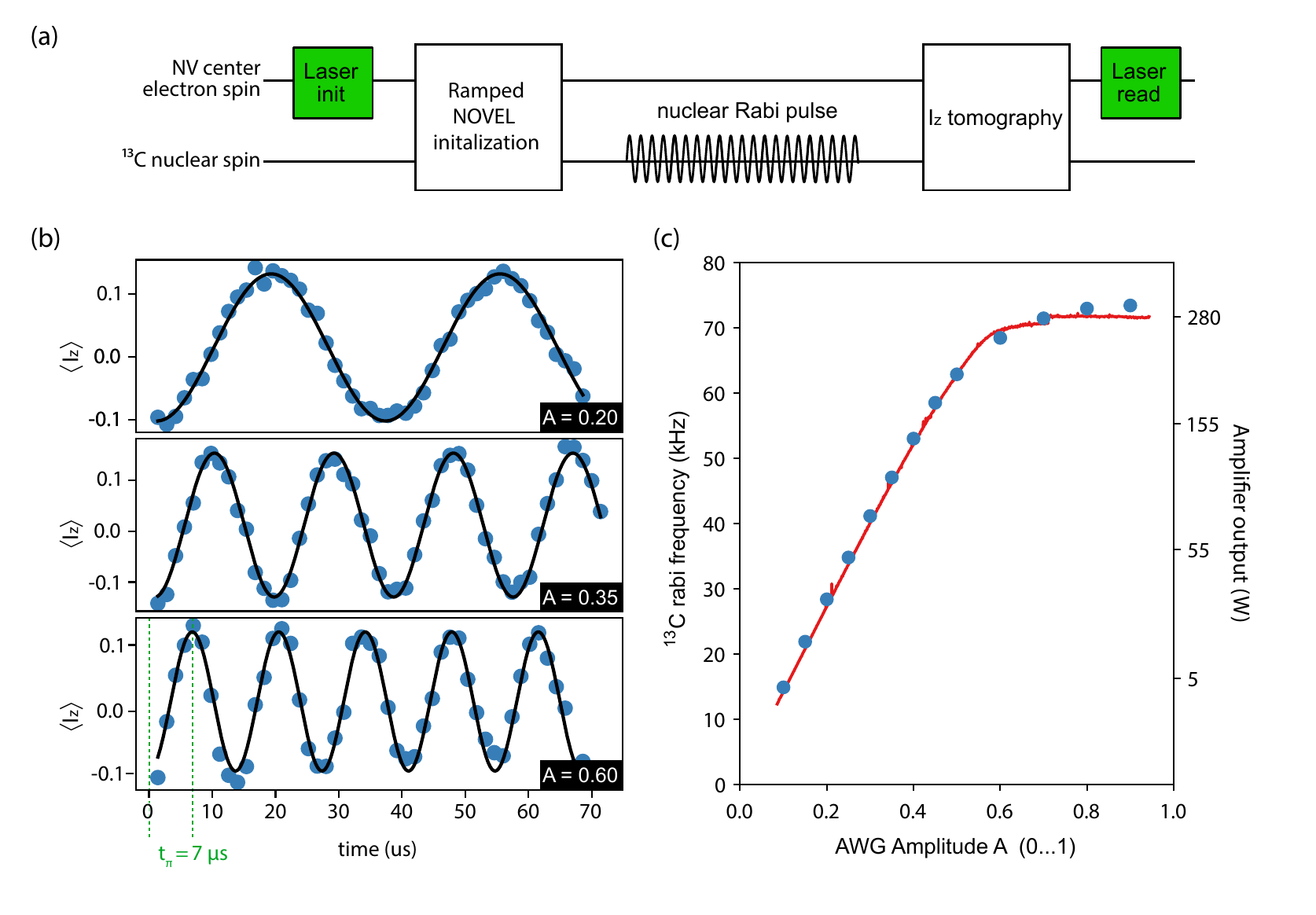}
	\caption{
	Demonstration of fast Rabi oscillations on a single \C nuclear spin.
(a) Experimental pulse sequence.  We initialize the nuclear spin by polarization transfer from the NV center using an amplitude-ramped NOVEL scheme \cite{henstra1988,can2017,cujia19}.  Subsequently, we apply a Rabi pulse of varying duration $t$ to the coil. The pulse is tuned into resonance with the \C Larmor frequency {at $2.13\unit{MHz}$}.  We detect the $I_z$ projection of the nuclear spin via the NV center using the state tomography protocol reported in Ref. \cite{taminiau14}.  A bias field of $199.4\unit{mT}$ is applied along the symmetry axis of the NV center.
(b) Examples for nuclear Rabi oscillations recorded at AWG amplitudes of $0.2$, $0.35$ and $0.6$.  Blue dots are the data and black lines are fits to sinusoids.
(c) Rabi frequency plotted as a function of relative AWG pulse amplitude (blue data points). Above an AWG amplitude of approximately 0.6, the amplifier saturates resulting in a maximum Rabi frequency of $74\unit{kHz}$.  The output power of the amplifier (red) is shown on the right scale.
	}
	\label{fig:rabi}
\end{figure*}

To examine the capability of our microcoil system for driving fast nuclear spin manipulation, we carry out a set of Rabi nutation experiments on a single $^{13}\mathrm{C}$ nuclear spin in the vicinity of the NV center.  We choose a \C that is sufficiently close to the NV spin to allow for a coherent coupling \cite{taminiau14}, yet distant enough to avoid significant hyperfine enhancement of the nuclear Rabi frequency \cite{sangtawesin16}. Our pulse sequence, sketched in Fig. \ref{fig:rabi}(a), consists of three steps: (i) Polarization of the \C nuclear spin, (ii) nuclear Rabi pulse of variable duration $t$, and (iii) detection of the $\braket{I_z}$ projection via nuclear state tomography.

Fig. \ref{fig:rabi}(b) shows three Rabi oscillation measurements for low, medium and high RF drive amplitudes.  We obtain the Rabi traces by repeating the protocol shown in Fig. \ref{fig:rabi}(a) and plotting $\braket{I_z}$ as a function of the nuclear pulse duration $t$.  We then extract the corresponding Rabi frequencies by fitting a simple sinusoid to the oscillation. Fig. \ref{fig:rabi}(c) shows the measured \C Rabi frequencies (blue dots) as a function of the normalized output amplitude of the AWG.  We also plot the peak output power of the amplifier (red curve).  The duty cycle of RF pulses in these experiments is between 4\% and 7\%. 

We observe a maximum Rabi frequency of $74\unit{kHz}$ at amplifier saturation ($\sim 280\unit{W}$ output power).  The high Rabi frequency permits nuclear $\pi/2$ rotations in $3.4\unit{\us}$ and $\pi$ rotation in $6.8\unit{\us}$, respectively (Fig. \ref{fig:rabi}b).  Taking into account the hyperfine enhancement of the nuclear Rabi frequency, which we estimate to $<4\%$ for this \C based on the expressions in Ref. \onlinecite{sangtawesin16}, the true Rabi frequency without enhancement is approximately $71\unit{kHz}$.  The corresponding magnetic field amplitude is $\sim 12.5\unit{mT}$.


\section{Conclusions}

In summary, we have designed a planar microcoil for efficient nuclear spin manipulation in the context of nanoscale NMR spectroscopy and solid-state quantum devices.  Our impedance-matched circuit has a large bandwidth of up to $\sim 20\unit{MHz}$ and provides $4.5\unit{mT/A}$ field at a separation of $1\unit{mm}$.  Higher bandwidths, larger fields or larger separations can be realized by adjusting winding numbers and coil radius.  Using an optimized heat removal structure, we are able to apply pulse amplitudes of up to $280\unit{W}$ at a $5\%$ duty cycle while keeping the temperature increase of the coil below $2\unit{K}$.  We demonstrate \C Rabi frequencies exceeding $70\unit{kHz}$ and nuclear $\pi/2$ and $\pi$ rotations in $3.4\unit{\us}$ and $6.8\unit{\us}$, respectively.  If desired, Rabi frequencies could be further enhanced by reducing the separation between microcoil and sample and by further increasing the amplifier power.

The large bandwidth of our microcoil circuit is especially useful for applications in low-field NMR spectroscopy, and in particular, for nanoscale NMR experiments that do not rely on thermal (Boltzmann) polarization \cite{Herzog14}.
Because the coil circuit is broadband, multiple nuclear species can be excited simultaneously by adding pulse patterns in software before uploading them onto the AWG hardware.  This feature will greatly simplify heteronuclear NMR schemes including polarization transfer, heteronuclear decoupling and two-dimensional correlation methods.
In addition, high RF fields are critical for homonuclear decoupling of proton spins in solids \cite{waugh68}.  Taking the reduced microcoil transmission at higher frequencies into account, the extrapolated \H Rabi frequency is about $240\unit{kHz}$ for a \H NMR frequency of $8\unit{MHz}$.  This Rabi frequency is well above the dipolar coupling frequencies even for dense proton networks, suggesting that homonuclear decoupling will be very efficient for a wide range of solid samples.

\section*{Acknowledgements}
This work was supported by Swiss National Science Foundation (SNFS) Project Grant No. 200020\_175600, the National Center of Competence in Research in Quantum Science and Technology (NCCR QSIT), and the Advancing Science and TEchnology thRough dIamond Quantum Sensing (ASTERQIS) program, Grant No. 820394, of the European Commission.
We thank Alexander D\"app for support with the characterization measurements of the Rohde \& Schwarz amplifier and Nils Hauff for discussions of initial designs.

\section*{Data Availability Statement}
The data that support the findings of this study are available from the corresponding author upon reasonable request.


\begin{thebibliography}{31}
\expandafter\ifx\csname natexlab\endcsname\relax\def\natexlab#1{#1}\fi
\expandafter\ifx\csname bibnamefont\endcsname\relax
  \def\bibnamefont#1{#1}\fi
\expandafter\ifx\csname bibfnamefont\endcsname\relax
  \def\bibfnamefont#1{#1}\fi
\expandafter\ifx\csname citenamefont\endcsname\relax
  \def\citenamefont#1{#1}\fi
\expandafter\ifx\csname url\endcsname\relax
  \def\url#1{\texttt{#1}}\fi
\expandafter\ifx\csname urlprefix\endcsname\relax\def\urlprefix{URL }\fi
\providecommand{\bibinfo}[2]{#2}
\providecommand{\eprint}[2][]{\url{#2}}

\bibitem[{\citenamefont{Waugh et~al.}(1968)\citenamefont{Waugh, Huber, and
  Haeberlen}}]{waugh68}
\bibinfo{author}{\bibfnamefont{J.~S.} \bibnamefont{Waugh}},
  \bibinfo{author}{\bibfnamefont{L.~M.} \bibnamefont{Huber}}, \bibnamefont{and}
  \bibinfo{author}{\bibfnamefont{U.}~\bibnamefont{Haeberlen}},
  \bibinfo{journal}{Phys. Rev. Lett.} \textbf{\bibinfo{volume}{20}},
  \bibinfo{pages}{180} (\bibinfo{year}{1968}),
  \urlprefix\url{https://link.aps.org/doi/10.1103/PhysRevLett.20.180}.

\bibitem[{\citenamefont{Aue et~al.}(1976)\citenamefont{Aue, Bartholdi, and
  Ernst}}]{aue76}
\bibinfo{author}{\bibfnamefont{W.~P.} \bibnamefont{Aue}},
  \bibinfo{author}{\bibfnamefont{E.}~\bibnamefont{Bartholdi}},
  \bibnamefont{and} \bibinfo{author}{\bibfnamefont{R.~R.} \bibnamefont{Ernst}},
  \bibinfo{journal}{J. Chem. Phys.} \textbf{\bibinfo{volume}{64}},
  \bibinfo{pages}{2229} (\bibinfo{year}{1976}).

\bibitem[{\citenamefont{Yamauchi et~al.}(2004)\citenamefont{Yamauchi, Janssen,
  and Kentgens}}]{yamauchi04}
\bibinfo{author}{\bibfnamefont{K.}~\bibnamefont{Yamauchi}},
  \bibinfo{author}{\bibfnamefont{J.}~\bibnamefont{Janssen}}, \bibnamefont{and}
  \bibinfo{author}{\bibfnamefont{A.}~\bibnamefont{Kentgens}},
  \bibinfo{journal}{J. Mag. Res.} \textbf{\bibinfo{volume}{167}},
  \bibinfo{pages}{87} (\bibinfo{year}{2004}),
  \urlprefix\url{http://www.sciencedirect.com/science/article/pii/S1090780703004245}.

\bibitem[{\citenamefont{Webb}(2013)}]{webb13}
\bibinfo{author}{\bibfnamefont{A.}~\bibnamefont{Webb}},
  \bibinfo{journal}{Journal of Magnetic Resonance}
  \textbf{\bibinfo{volume}{229}}, \bibinfo{pages}{55} (\bibinfo{year}{2013}),
  \urlprefix\url{http://www.sciencedirect.com/science/article/pii/S1090780712003187}.

\bibitem[{\citenamefont{Staudacher et~al.}(2013)\citenamefont{Staudacher, Shi,
  Pezzagna, Meijer, Du, Meriles, Reinhard, and Wrachtrup}}]{staudacher13}
\bibinfo{author}{\bibfnamefont{T.}~\bibnamefont{Staudacher}},
  \bibinfo{author}{\bibfnamefont{F.}~\bibnamefont{Shi}},
  \bibinfo{author}{\bibfnamefont{S.}~\bibnamefont{Pezzagna}},
  \bibinfo{author}{\bibfnamefont{J.}~\bibnamefont{Meijer}},
  \bibinfo{author}{\bibfnamefont{J.}~\bibnamefont{Du}},
  \bibinfo{author}{\bibfnamefont{C.~A.} \bibnamefont{Meriles}},
  \bibinfo{author}{\bibfnamefont{F.}~\bibnamefont{Reinhard}}, \bibnamefont{and}
  \bibinfo{author}{\bibfnamefont{J.}~\bibnamefont{Wrachtrup}},
  \bibinfo{journal}{Science} \textbf{\bibinfo{volume}{339}},
  \bibinfo{pages}{561} (\bibinfo{year}{2013}).

\bibitem[{\citenamefont{Mamin et~al.}(2013)\citenamefont{Mamin, Kim, Sherwood,
  Rettner, Ohno, Awschalom, and Rugar}}]{mamin13}
\bibinfo{author}{\bibfnamefont{H.~J.} \bibnamefont{Mamin}},
  \bibinfo{author}{\bibfnamefont{M.}~\bibnamefont{Kim}},
  \bibinfo{author}{\bibfnamefont{M.~H.} \bibnamefont{Sherwood}},
  \bibinfo{author}{\bibfnamefont{C.~T.} \bibnamefont{Rettner}},
  \bibinfo{author}{\bibfnamefont{K.}~\bibnamefont{Ohno}},
  \bibinfo{author}{\bibfnamefont{D.~D.} \bibnamefont{Awschalom}},
  \bibnamefont{and} \bibinfo{author}{\bibfnamefont{D.}~\bibnamefont{Rugar}},
  \bibinfo{journal}{Science} \textbf{\bibinfo{volume}{339}},
  \bibinfo{pages}{557} (\bibinfo{year}{2013}).

\bibitem[{\citenamefont{Glenn et~al.}(2018)\citenamefont{Glenn, Bucher, Lee,
  Lukin, Park, and Walsworth}}]{glenn18}
\bibinfo{author}{\bibfnamefont{D.~R.} \bibnamefont{Glenn}},
  \bibinfo{author}{\bibfnamefont{D.~B.} \bibnamefont{Bucher}},
  \bibinfo{author}{\bibfnamefont{J.}~\bibnamefont{Lee}},
  \bibinfo{author}{\bibfnamefont{M.~D.} \bibnamefont{Lukin}},
  \bibinfo{author}{\bibfnamefont{H.}~\bibnamefont{Park}}, \bibnamefont{and}
  \bibinfo{author}{\bibfnamefont{R.~L.} \bibnamefont{Walsworth}},
  \bibinfo{journal}{Nature} \textbf{\bibinfo{volume}{555}},
  \bibinfo{pages}{351} (\bibinfo{year}{2018}),
  \urlprefix\url{http://dx.doi.org/10.1038/nature25781}.

\bibitem[{\citenamefont{Smits et~al.}(2019)\citenamefont{Smits, Damron,
  Kehayias, Mcdowell, Mosavian, Fescenko, Ristoff, Laraoui, Jarmola, and
  Acosta}}]{smits19}
\bibinfo{author}{\bibfnamefont{J.}~\bibnamefont{Smits}},
  \bibinfo{author}{\bibfnamefont{J.~T.} \bibnamefont{Damron}},
  \bibinfo{author}{\bibfnamefont{P.}~\bibnamefont{Kehayias}},
  \bibinfo{author}{\bibfnamefont{A.~F.} \bibnamefont{Mcdowell}},
  \bibinfo{author}{\bibfnamefont{N.}~\bibnamefont{Mosavian}},
  \bibinfo{author}{\bibfnamefont{I.}~\bibnamefont{Fescenko}},
  \bibinfo{author}{\bibfnamefont{N.}~\bibnamefont{Ristoff}},
  \bibinfo{author}{\bibfnamefont{A.}~\bibnamefont{Laraoui}},
  \bibinfo{author}{\bibfnamefont{A.}~\bibnamefont{Jarmola}}, \bibnamefont{and}
  \bibinfo{author}{\bibfnamefont{V.~M.} \bibnamefont{Acosta}},
  \bibinfo{journal}{Sci Adv} \textbf{\bibinfo{volume}{5}},
  \bibinfo{pages}{eaaw7895} (\bibinfo{year}{2019}).

\bibitem[{\citenamefont{Ajoy et~al.}(2015)\citenamefont{Ajoy, Bissbort, Lukin,
  Walsworth, and Cappellaro}}]{ajoy15}
\bibinfo{author}{\bibfnamefont{A.}~\bibnamefont{Ajoy}},
  \bibinfo{author}{\bibfnamefont{U.}~\bibnamefont{Bissbort}},
  \bibinfo{author}{\bibfnamefont{M.~D.} \bibnamefont{Lukin}},
  \bibinfo{author}{\bibfnamefont{R.~L.} \bibnamefont{Walsworth}},
  \bibnamefont{and}
  \bibinfo{author}{\bibfnamefont{P.}~\bibnamefont{Cappellaro}},
  \bibinfo{journal}{Phys. Rev. X} \textbf{\bibinfo{volume}{5}},
  \bibinfo{pages}{011001} (\bibinfo{year}{2015}).

\bibitem[{\citenamefont{Wrachtrup and Finkler}(2016)}]{wrachtrup16}
\bibinfo{author}{\bibfnamefont{J.}~\bibnamefont{Wrachtrup}} \bibnamefont{and}
  \bibinfo{author}{\bibfnamefont{A.}~\bibnamefont{Finkler}},
  \bibinfo{journal}{J. Magn. Reson.} \textbf{\bibinfo{volume}{269}},
  \bibinfo{pages}{225} (\bibinfo{year}{2016}).

\bibitem[{\citenamefont{Schwartz et~al.}(2017)\citenamefont{Schwartz, Rosskopf,
  Schmitt, Tratzmiller, Chen, McGuinness, Jelezko, and Plenio}}]{schwartz17}
\bibinfo{author}{\bibfnamefont{I.}~\bibnamefont{Schwartz}},
  \bibinfo{author}{\bibfnamefont{J.}~\bibnamefont{Rosskopf}},
  \bibinfo{author}{\bibfnamefont{S.}~\bibnamefont{Schmitt}},
  \bibinfo{author}{\bibfnamefont{B.}~\bibnamefont{Tratzmiller}},
  \bibinfo{author}{\bibfnamefont{Q.}~\bibnamefont{Chen}},
  \bibinfo{author}{\bibfnamefont{L.~P.} \bibnamefont{McGuinness}},
  \bibinfo{author}{\bibfnamefont{F.}~\bibnamefont{Jelezko}}, \bibnamefont{and}
  \bibinfo{author}{\bibfnamefont{M.~B.} \bibnamefont{Plenio}},
  \bibinfo{journal}{arXiv:1706.07134}  (\bibinfo{year}{2017}),
  \urlprefix\url{https://arxiv.org/abs/1706.07134}.

\bibitem[{\citenamefont{Dutt et~al.}(2007)\citenamefont{Dutt, Childress, Jiang,
  Togan, Maze, Jelezko, Zibrov, Hemmer, and Lukin}}]{dutt07}
\bibinfo{author}{\bibfnamefont{M.~V.~G.} \bibnamefont{Dutt}},
  \bibinfo{author}{\bibfnamefont{L.}~\bibnamefont{Childress}},
  \bibinfo{author}{\bibfnamefont{L.}~\bibnamefont{Jiang}},
  \bibinfo{author}{\bibfnamefont{E.}~\bibnamefont{Togan}},
  \bibinfo{author}{\bibfnamefont{J.}~\bibnamefont{Maze}},
  \bibinfo{author}{\bibfnamefont{F.}~\bibnamefont{Jelezko}},
  \bibinfo{author}{\bibfnamefont{A.~S.} \bibnamefont{Zibrov}},
  \bibinfo{author}{\bibfnamefont{P.~R.} \bibnamefont{Hemmer}},
  \bibnamefont{and} \bibinfo{author}{\bibfnamefont{M.~D.} \bibnamefont{Lukin}},
  \bibinfo{journal}{Science} \textbf{\bibinfo{volume}{316}},
  \bibinfo{eid}{1312} (pages~\bibinfo{numpages}{5}) (\bibinfo{year}{2007}),
  \urlprefix\url{http://dx.doi.org/10.1126/science.1139831}.

\bibitem[{\citenamefont{Pla et~al.}(2013)\citenamefont{Pla, Tan, Dehollain,
  Lim, Morton, Zwanenburg, Jamieson, Dzurak, and Morello}}]{pla13}
\bibinfo{author}{\bibfnamefont{J.~J.} \bibnamefont{Pla}},
  \bibinfo{author}{\bibfnamefont{K.~Y.} \bibnamefont{Tan}},
  \bibinfo{author}{\bibfnamefont{J.~P.} \bibnamefont{Dehollain}},
  \bibinfo{author}{\bibfnamefont{W.~H.} \bibnamefont{Lim}},
  \bibinfo{author}{\bibfnamefont{J.~J.~L.} \bibnamefont{Morton}},
  \bibinfo{author}{\bibfnamefont{F.~A.} \bibnamefont{Zwanenburg}},
  \bibinfo{author}{\bibfnamefont{D.~N.} \bibnamefont{Jamieson}},
  \bibinfo{author}{\bibfnamefont{A.~S.} \bibnamefont{Dzurak}},
  \bibnamefont{and} \bibinfo{author}{\bibfnamefont{A.}~\bibnamefont{Morello}},
  \bibinfo{journal}{Nature} \textbf{\bibinfo{volume}{496}}
  (\bibinfo{year}{2013}), \urlprefix\url{https://doi.org/10.1038/nature12011}.

\bibitem[{\citenamefont{Taminiau et~al.}(2014)\citenamefont{Taminiau, Cramer,
  van~der Sar, Dobrovitski, and Hanson}}]{taminiau14}
\bibinfo{author}{\bibfnamefont{T.~H.} \bibnamefont{Taminiau}},
  \bibinfo{author}{\bibfnamefont{J.}~\bibnamefont{Cramer}},
  \bibinfo{author}{\bibfnamefont{T.}~\bibnamefont{van~der Sar}},
  \bibinfo{author}{\bibfnamefont{V.~V.} \bibnamefont{Dobrovitski}},
  \bibnamefont{and} \bibinfo{author}{\bibfnamefont{R.}~\bibnamefont{Hanson}},
  \bibinfo{journal}{Nature Nano.} \textbf{\bibinfo{volume}{9}},
  \bibinfo{pages}{171} (\bibinfo{year}{2014}).

\bibitem[{\citenamefont{Bradley et~al.}(2019)\citenamefont{Bradley, Randall,
  Abobeih, Berrevoets, Degen, Bakker, Markham, Twitchen, and
  Taminiau}}]{bradley19}
\bibinfo{author}{\bibfnamefont{C.}~\bibnamefont{Bradley}},
  \bibinfo{author}{\bibfnamefont{J.}~\bibnamefont{Randall}},
  \bibinfo{author}{\bibfnamefont{M.}~\bibnamefont{Abobeih}},
  \bibinfo{author}{\bibfnamefont{R.}~\bibnamefont{Berrevoets}},
  \bibinfo{author}{\bibfnamefont{M.}~\bibnamefont{Degen}},
  \bibinfo{author}{\bibfnamefont{M.}~\bibnamefont{Bakker}},
  \bibinfo{author}{\bibfnamefont{M.}~\bibnamefont{Markham}},
  \bibinfo{author}{\bibfnamefont{D.}~\bibnamefont{Twitchen}}, \bibnamefont{and}
  \bibinfo{author}{\bibfnamefont{T.}~\bibnamefont{Taminiau}},
  \bibinfo{journal}{Phys. Rev. X} \textbf{\bibinfo{volume}{9}},
  \bibinfo{pages}{031045} (\bibinfo{year}{2019}),
  \urlprefix\url{https://link.aps.org/doi/10.1103/PhysRevX.9.031045}.

\bibitem[{\citenamefont{Morton et~al.}(2008)\citenamefont{Morton, Tyryshkin,
  Brown, Shankar, Lovett, Ardavan, Schenkel, Haller, Ager, and
  Lyon}}]{morton08}
\bibinfo{author}{\bibfnamefont{J.~J.~L.} \bibnamefont{Morton}},
  \bibinfo{author}{\bibfnamefont{A.~M.} \bibnamefont{Tyryshkin}},
  \bibinfo{author}{\bibfnamefont{R.~M.} \bibnamefont{Brown}},
  \bibinfo{author}{\bibfnamefont{S.}~\bibnamefont{Shankar}},
  \bibinfo{author}{\bibfnamefont{B.~W.} \bibnamefont{Lovett}},
  \bibinfo{author}{\bibfnamefont{A.}~\bibnamefont{Ardavan}},
  \bibinfo{author}{\bibfnamefont{T.}~\bibnamefont{Schenkel}},
  \bibinfo{author}{\bibfnamefont{E.~E.} \bibnamefont{Haller}},
  \bibinfo{author}{\bibfnamefont{J.~W.} \bibnamefont{Ager}}, \bibnamefont{and}
  \bibinfo{author}{\bibfnamefont{S.~A.} \bibnamefont{Lyon}},
  \bibinfo{journal}{Nature} \textbf{\bibinfo{volume}{455}}
  (\bibinfo{year}{2008}), \urlprefix\url{https://doi.org/10.1038/nature07295}.

\bibitem[{\citenamefont{Pfender et~al.}(2017)\citenamefont{Pfender, Aslam,
  Sumiya, Onoda, Neumann, Isoya, Meriles, and Wrachtrup}}]{pfender17}
\bibinfo{author}{\bibfnamefont{M.}~\bibnamefont{Pfender}},
  \bibinfo{author}{\bibfnamefont{N.}~\bibnamefont{Aslam}},
  \bibinfo{author}{\bibfnamefont{H.}~\bibnamefont{Sumiya}},
  \bibinfo{author}{\bibfnamefont{S.}~\bibnamefont{Onoda}},
  \bibinfo{author}{\bibfnamefont{P.}~\bibnamefont{Neumann}},
  \bibinfo{author}{\bibfnamefont{J.}~\bibnamefont{Isoya}},
  \bibinfo{author}{\bibfnamefont{C.~A.} \bibnamefont{Meriles}},
  \bibnamefont{and}
  \bibinfo{author}{\bibfnamefont{J.}~\bibnamefont{Wrachtrup}},
  \bibinfo{journal}{Nature Communications} \textbf{\bibinfo{volume}{8}},
  \bibinfo{pages}{834} (\bibinfo{year}{2017}),
  \urlprefix\url{https://doi.org/10.1038/s41467-017-00964-z}.

\bibitem[{\citenamefont{Rosskopf et~al.}(2017)\citenamefont{Rosskopf, Zopes,
  Boss, and Degen}}]{rosskopf17}
\bibinfo{author}{\bibfnamefont{T.}~\bibnamefont{Rosskopf}},
  \bibinfo{author}{\bibfnamefont{J.}~\bibnamefont{Zopes}},
  \bibinfo{author}{\bibfnamefont{J.~M.} \bibnamefont{Boss}}, \bibnamefont{and}
  \bibinfo{author}{\bibfnamefont{C.~L.} \bibnamefont{Degen}},
  \bibinfo{journal}{NPJ Quantum Information} \textbf{\bibinfo{volume}{3}},
  \bibinfo{pages}{33} (\bibinfo{year}{2017}),
  \urlprefix\url{http://www.nature.com/articles/s41534-017-0030-6}.

\bibitem[{\citenamefont{Lovchinsky et~al.}(2016)\citenamefont{Lovchinsky,
  Sushkov, Urbach, de~Leon, Choi, de~Greve, Evans, Gertner, Bersin, Muller
  et~al.}}]{lovchinsky16}
\bibinfo{author}{\bibfnamefont{I.}~\bibnamefont{Lovchinsky}},
  \bibinfo{author}{\bibfnamefont{A.~O.} \bibnamefont{Sushkov}},
  \bibinfo{author}{\bibfnamefont{E.}~\bibnamefont{Urbach}},
  \bibinfo{author}{\bibfnamefont{N.~P.} \bibnamefont{de~Leon}},
  \bibinfo{author}{\bibfnamefont{S.}~\bibnamefont{Choi}},
  \bibinfo{author}{\bibfnamefont{K.}~\bibnamefont{de~Greve}},
  \bibinfo{author}{\bibfnamefont{R.}~\bibnamefont{Evans}},
  \bibinfo{author}{\bibfnamefont{R.}~\bibnamefont{Gertner}},
  \bibinfo{author}{\bibfnamefont{E.}~\bibnamefont{Bersin}},
  \bibinfo{author}{\bibfnamefont{C.}~\bibnamefont{Muller}},
  \bibnamefont{et~al.}, \bibinfo{journal}{Science}
  \textbf{\bibinfo{volume}{351}}, \bibinfo{pages}{836} (\bibinfo{year}{2016}),
  \urlprefix\url{http://science.sciencemag.org/content/351/6275/836.abstract}.

\bibitem[{\citenamefont{Zopes et~al.}(2018)\citenamefont{Zopes, Herb, Cujia,
  and Degen}}]{zopes18prl}
\bibinfo{author}{\bibfnamefont{J.}~\bibnamefont{Zopes}},
  \bibinfo{author}{\bibfnamefont{K.}~\bibnamefont{Herb}},
  \bibinfo{author}{\bibfnamefont{K.~S.} \bibnamefont{Cujia}}, \bibnamefont{and}
  \bibinfo{author}{\bibfnamefont{C.~L.} \bibnamefont{Degen}},
  \bibinfo{journal}{Phys. Rev. Lett.} \textbf{\bibinfo{volume}{121}},
  \bibinfo{pages}{170801} (\bibinfo{year}{2018}),
  \urlprefix\url{https://link.aps.org/doi/10.1103/PhysRevLett.121.170801}.

\bibitem[{\citenamefont{Dreau et~al.}(2011)\citenamefont{Dreau, Lesik, Rondin,
  Spinicelli, Arcizet, Roch, and Jacques}}]{dreau11}
\bibinfo{author}{\bibfnamefont{A.}~\bibnamefont{Dreau}},
  \bibinfo{author}{\bibfnamefont{M.}~\bibnamefont{Lesik}},
  \bibinfo{author}{\bibfnamefont{L.}~\bibnamefont{Rondin}},
  \bibinfo{author}{\bibfnamefont{P.}~\bibnamefont{Spinicelli}},
  \bibinfo{author}{\bibfnamefont{O.}~\bibnamefont{Arcizet}},
  \bibinfo{author}{\bibfnamefont{J.~F.} \bibnamefont{Roch}}, \bibnamefont{and}
  \bibinfo{author}{\bibfnamefont{V.}~\bibnamefont{Jacques}},
  \bibinfo{journal}{Phys. Rev. B} \textbf{\bibinfo{volume}{84}},
  \bibinfo{pages}{195204} (\bibinfo{year}{2011}).

\bibitem[{\citenamefont{Bloch and Siegert}(1940)}]{bloch40}
\bibinfo{author}{\bibfnamefont{F.}~\bibnamefont{Bloch}} \bibnamefont{and}
  \bibinfo{author}{\bibfnamefont{A.}~\bibnamefont{Siegert}},
  \bibinfo{journal}{Phys. Rev.} \textbf{\bibinfo{volume}{57}},
  \bibinfo{pages}{522} (\bibinfo{year}{1940}),
  \urlprefix\url{https://link.aps.org/doi/10.1103/PhysRev.57.522}.

\bibitem[{\citenamefont{Sacolick et~al.}(2010)\citenamefont{Sacolick,
  Wiesinger, Hancu, and Vogel}}]{sacolick10}
\bibinfo{author}{\bibfnamefont{L.~I.} \bibnamefont{Sacolick}},
  \bibinfo{author}{\bibfnamefont{F.}~\bibnamefont{Wiesinger}},
  \bibinfo{author}{\bibfnamefont{I.}~\bibnamefont{Hancu}}, \bibnamefont{and}
  \bibinfo{author}{\bibfnamefont{M.~W.} \bibnamefont{Vogel}},
  \bibinfo{journal}{Magnetic Resonance in Medicine}
  \textbf{\bibinfo{volume}{63}}, \bibinfo{pages}{1315} (\bibinfo{year}{2010}),
  \urlprefix\url{https://doi.org/10.1002/mrm.22357}.

\bibitem[{\citenamefont{Silveri et~al.}(2017)\citenamefont{Silveri, Tuorila,
  Thuneberg, and Paraoanu}}]{Silveri2017}
\bibinfo{author}{\bibfnamefont{M.~P.} \bibnamefont{Silveri}},
  \bibinfo{author}{\bibfnamefont{J.~A.} \bibnamefont{Tuorila}},
  \bibinfo{author}{\bibfnamefont{E.~V.} \bibnamefont{Thuneberg}},
  \bibnamefont{and} \bibinfo{author}{\bibfnamefont{G.~S.}
  \bibnamefont{Paraoanu}}, \bibinfo{journal}{Reports on Progress in Physics}
  \textbf{\bibinfo{volume}{80}}, \bibinfo{pages}{056002}
  (\bibinfo{year}{2017}).

\bibitem[{\citenamefont{Henstra et~al.}(1988)\citenamefont{Henstra, Dirksen,
  Schmidt, and Wenckebach}}]{henstra1988}
\bibinfo{author}{\bibfnamefont{A.}~\bibnamefont{Henstra}},
  \bibinfo{author}{\bibfnamefont{P.}~\bibnamefont{Dirksen}},
  \bibinfo{author}{\bibfnamefont{J.}~\bibnamefont{Schmidt}}, \bibnamefont{and}
  \bibinfo{author}{\bibfnamefont{W.}~\bibnamefont{Wenckebach}},
  \bibinfo{journal}{Journal of Magnetic Resonance (1969)}
  \textbf{\bibinfo{volume}{77}}, \bibinfo{pages}{389 } (\bibinfo{year}{1988}),
  ISSN \bibinfo{issn}{0022-2364}.

\bibitem[{\citenamefont{Can et~al.}(2017)\citenamefont{Can, Weber, Walish,
  Swager, and Griffin}}]{can2017}
\bibinfo{author}{\bibfnamefont{T.~V.} \bibnamefont{Can}},
  \bibinfo{author}{\bibfnamefont{R.~T.} \bibnamefont{Weber}},
  \bibinfo{author}{\bibfnamefont{J.~J.} \bibnamefont{Walish}},
  \bibinfo{author}{\bibfnamefont{T.~M.} \bibnamefont{Swager}},
  \bibnamefont{and} \bibinfo{author}{\bibfnamefont{R.~G.}
  \bibnamefont{Griffin}}, \bibinfo{journal}{The Journal of Chemical Physics}
  \textbf{\bibinfo{volume}{146}}, \bibinfo{pages}{154204}
  (\bibinfo{year}{2017}).

\bibitem[{\citenamefont{Cujia et~al.}(2019)\citenamefont{Cujia, Boss, Herb,
  Zopes, and Degen}}]{cujia19}
\bibinfo{author}{\bibfnamefont{K.~S.} \bibnamefont{Cujia}},
  \bibinfo{author}{\bibfnamefont{J.~M.} \bibnamefont{Boss}},
  \bibinfo{author}{\bibfnamefont{K.}~\bibnamefont{Herb}},
  \bibinfo{author}{\bibfnamefont{J.}~\bibnamefont{Zopes}}, \bibnamefont{and}
  \bibinfo{author}{\bibfnamefont{C.~L.} \bibnamefont{Degen}},
  \bibinfo{journal}{Nature} \textbf{\bibinfo{volume}{571}},
  \bibinfo{pages}{1476} (\bibinfo{year}{2019}),
  \urlprefix\url{https://doi.org/10.1038/s41586-019-1334-9}.

\bibitem[{\citenamefont{Sangtawesin et~al.}(2016)\citenamefont{Sangtawesin,
  McLellan, Myers, Jayich, Awschalom, and Petta}}]{sangtawesin16}
\bibinfo{author}{\bibfnamefont{S.}~\bibnamefont{Sangtawesin}},
  \bibinfo{author}{\bibfnamefont{C.~A.} \bibnamefont{McLellan}},
  \bibinfo{author}{\bibfnamefont{B.~A.} \bibnamefont{Myers}},
  \bibinfo{author}{\bibfnamefont{A.~C.~B.} \bibnamefont{Jayich}},
  \bibinfo{author}{\bibfnamefont{D.~D.} \bibnamefont{Awschalom}},
  \bibnamefont{and} \bibinfo{author}{\bibfnamefont{J.~R.} \bibnamefont{Petta}},
  \bibinfo{journal}{New Journal of Physics} \textbf{\bibinfo{volume}{18}},
  \bibinfo{pages}{083016} (\bibinfo{year}{2016}),
  \urlprefix\url{https://doi.org/10.1088%2F1367-2630%2F18%2F8%2F083016}.

\bibitem[{\citenamefont{Herzog et~al.}(2014)\citenamefont{Herzog, Cadeddu, Xue,
  Peddibhotla, and Poggio}}]{Herzog14}
\bibinfo{author}{\bibfnamefont{B.~E.} \bibnamefont{Herzog}},
  \bibinfo{author}{\bibfnamefont{D.}~\bibnamefont{Cadeddu}},
  \bibinfo{author}{\bibfnamefont{F.}~\bibnamefont{Xue}},
  \bibinfo{author}{\bibfnamefont{P.}~\bibnamefont{Peddibhotla}},
  \bibnamefont{and} \bibinfo{author}{\bibfnamefont{M.}~\bibnamefont{Poggio}},
  \bibinfo{journal}{Appl. Phys. Lett.} \textbf{\bibinfo{volume}{105}},
  \bibinfo{pages}{043112} (\bibinfo{year}{2014}).

\bibitem[{\citenamefont{Giacoletto}(1977)}]{Electronics_handbook}
\bibinfo{author}{\bibfnamefont{L.~J.} \bibnamefont{Giacoletto}},
  \emph{\bibinfo{title}{Electronics Designer's Handbook}}
  (\bibinfo{publisher}{McGraw-Hill}, \bibinfo{year}{1977}), ISBN
  \bibinfo{isbn}{0070231494}.

\bibitem[{\citenamefont{Mehring}(1983)}]{Mehring1983}
\bibinfo{author}{\bibfnamefont{M.}~\bibnamefont{Mehring}},
  \emph{\bibinfo{title}{Principles of High Resolution NMR in Solids}}
  (\bibinfo{publisher}{Springe}, \bibinfo{address}{Berlin Heidelberg},
  \bibinfo{year}{1983}), ISBN \bibinfo{isbn}{978-3-642-68756-3}.

\end{thebibliography}

\appendix
\section{Empirical expressions for coil inductance and skin effect}

\subsection*{Inductance}
The geometry-dependent factors $G_1$ and $G_2$ of Eq. (\ref{eq:L}) are approximately given by
\begin{subequations}
\begin{equation}
\label{corr_fact1}
\begin{split}
G_1 &= \log\left(\frac{8 \bar{R}}{R_2-R_1}\right) - \frac{1}{2} \\ &+  \frac{1}{24} \left(\frac{R_2-R_1}{2 \bar{R}}\right)^2 \left[\log\left(\frac{8 \bar{R}}{R_2-R_1}\right)+3.583\right]
\end{split}
\end{equation}
and
\begin{equation}
\label{corr_fact2}
G_2 = \left[ 1+0.97\cdot\left(\frac{R_2-R_1}{2 \bar{R}}\right)^{0.62} \cdot \frac{N_1 d_\mathrm{w}}{R_2-R_1} \right]^{-1}
\end{equation}
\end{subequations}
according to Ref. \cite[p. 3-34 ff.]{Electronics_handbook}.  Here, $R_1$ is the inner radius, $R_2 = R_1 + N_2 \dw/2$ the outer radius, and $\dw$ is the wire diameter.  $N_1$ the number of layers and $N_2$ the number of windings per layer.

\subsection*{Skin effect}

The correction factor $\alpha$ in Eq. (\ref{eq:P}), which describes the increase in resistance due to the skin effect, is approximately given by
\begin{equation}
\label{eq:skin}
\alpha^{-1}(x) = \frac{2}{x^2} \left[x-1+\mathrm{e}^{-x}\right]
\end{equation}
where $x= \sqrt{f}\dw/2\delta$.  Here $f$ is the frequency, $\dw$ the wire diameter, and $\delta = 0.06611 \unit{\sqrt{Hz}\,m}$ the skin depth parameter for copper.  The correction factor $\alpha$ is $1.27$ at $1\unit{MHz}$, $1.48$ at $3\unit{MHz}$ and $1.93$ at $10\unit{MHz}$.

\section{Bloch-Siegert shift for spin $S=1$ system:}

The Bloch-Siegert shift for a spin $S=1$ system is given by (see also Ref. \onlinecite[p. 316 ff., adapted for a spin $S=1$ system]{Mehring1983}):
\begin{equation}
\label{eq:dfexact}
\begin{split}
\df &= \left( \frac{2}{f_0^{(1)}+\frf} +  \frac{2}{f_0^{(1)}-\frf}
\right. \\
 &+ \left.\frac{1}{f_0^{(2)}+\frf} +  \frac{1}{f_0^{(2)}-\frf} \right)
\frac{\left(\ye B_\perp\right)^2}{8}  \text{,}
\end{split}
\end{equation}
where $f_0^{(1)}$ denotes the frequency of the observed transition under no RF irradiation, $f_0^{(2)}$ the frequency of the unobserved transition under no RF irradiation, $\frf$ the RF frequency applied to the coil, and $\ye=28.024\unit{MHz/mT}$.  In our experiment, $f_0^{(1)}$ is the lower frequency ($\ms=0$ to $\ms=-1$) and $f_0^{(2)}$ is the higher frequency ($\ms=0$ to $\ms=1$) transition.  For low RF frequencies $\frf \ll f_0^{(1)},f_0^{(2)}$, the above expression simplifies to
\begin{equation}
\df \approx \frac{f_0^{(1)}+2f_0^{(2)}}{4f_0^{(1)}f_0^{(2)}} \ye^2 \Bperp^2
\tag{\ref{eq:df}} 
\end{equation}

\end{document}